\documentstyle[11pt,epsf]{article}
\newlength{\bredde}
\def\slash#1{\settowidth{\bredde}{$#1$}\ifmmode\,\raisebox{.15ex}{/}
\hspace*{-\bredde} #1\else$\,\raisebox{.15ex}{/}\hspace*{-\bredde} #1$\fi}
\newcommand{\beq}{\begin{equation}}
\newcommand{\eeq}{\end{equation}}
\newcommand{\bea}{\begin{eqnarray}}
\newcommand{\eea}{\end{eqnarray}}
\newcommand{\ba}{\begin{array}}
\newcommand{\ea}{\end{array}}
\newcommand{\noi}{\vspace{12pt}\noindent}
\def\to{\rightarrow}
\def\la{\lambda}
\def\l{\lambda}
\def\a{\alpha}
\renewcommand{\thefootnote}{ }

\begin{document}
\topmargin -1.4cm
\oddsidemargin -0.8cm
\evensidemargin -0.8cm
\title{\Large{{\bf 
Universal Spectral Correlators and Massive Dirac Operators}}}

\vspace{1.5cm}

\author{~\\~\\
{\sc Poul H. Damgaard}\\
The Niels Bohr Institute\\ Blegdamsvej 17\\ DK-2100 Copenhagen\\
Denmark\\~\\and\\~\\
{\sc Shinsuke M. Nishigaki}\\
Institute for Theoretical Physics\\University of California\\
Santa Barbara, CA 93106\\USA
}
 
\maketitle
\vfill
\begin{abstract} 
We derive the large-$N$ spectral correlators of
complex matrix ensembles with weights that in the context
of Dirac spectra correspond to $N_f$ massive fermions, and prove that
the results are universal in the appropriate scaling limits. The
resulting microscopic spectral densities satisfy exact spectral sum
rules of massive Dirac operators in QCD.
\end{abstract}
\vfill
\begin{flushleft}
NBI-HE-97-56 \\
NSF-ITP-97-138\\
hep-th/9711023
\end{flushleft}
\thispagestyle{empty}
\footnotetext{
e-mail addresses: {\tt phdamg@nbi.dk},\ {\tt shinsuke@itp.ucsb.edu}}

\newpage

\renewcommand{\thefootnote}{\fnsymbol{footnote}}
\setcounter{footnote}{0}
\setcounter{page}{1}
\section{Introduction}

There has recently been remarkable progress in the understanding of
chiral symmetry breaking in QCD, based on universality conjectures from
random matrix theory \cite{SV,V,VZ}. Central in this development is the
idea that the spectra of massless Dirac operators in gauge theories that have 
non-vanishing chiral condensates (as is the case when there is 
spontaneous chiral symmetry breaking) display universal large-volume 
scaling laws near the origin.
There is now mounting
evidence that this scenario is correct. First, the appropriate microscopic
spectral densities, derived from large-$N$ matrix ensembles with Gaussian
weights \cite{V,VZ}, have been shown to consistently reproduce the exact
spectral sum rules of Leutwyler and Smilga \cite{LS}. Second, the
microscopic spectral densities, and in fact all microscopic spectral
correlators, have been proven to be {\em universal} within the 
given classes of matrix model ensembles \cite{us}. Third, there is now
direct numerical support from Monte Carlo simulations of lattice gauge
theory (for the case of gauge group $SU(2)$ and quenched staggered fermions)
that the microscopic limit of lattice Dirac spectra is as predicted from
random matrix theory \cite{BBMSVW}.

\noi
In quenched Monte Carlo simulations it is possible to compute, for finite
volume, the lattice eigenvalues for massless quarks. In more realistic
lattice computations one will keep finite quark masses, and at most
consider mass rescalings towards smaller values as the volume is increased.
So for the practical purpose of comparison with lattice gauge
theory results, it is imperative that finite quark mass effects are
under control and understood. From a purely theoretical point of view 
there is also interest in the problem of finite masses. 
Just as one considers a microscopic
rescaling of eigenvalues $z$
it seems natural, from the Dirac eigenvalue equation,
to consider also the corresponding microscopic scaling limit in $m$
\cite{SV,JNZ}. In view of the above, one might hope that such a 
double-microscopic limit of the Dirac spectrum is universal as well, and
computable from random matrix theory. 

\noi
The starting point of the following computations are the suggested matrix
model ensembles that should be relevant for massive Dirac fermions 
\cite{V,VZ}. 
{}From this
we shall derive the pertinent spectral correlators (and in particular
the spectral densities themselves) for $N_f$ massive fermions in the 
double-microscopic limit where both eigenvalues $\lambda$ and masses $m$
are appropriately rescaled with $N$. As a straightforward by-product of the
analysis in ref.~\cite{us}, we shall prove that all of these 
double-microscopic spectral correlators are universal. We shall also
show that the resulting microscopic spectral densities are consistent
with exact spectral sum rules for massive Dirac operators in QCD and
QCD-like theories. Such spectral sum rules are normally derived for
the case of massless Dirac operators \cite{LS}, but the generalization
to massive Dirac operators is not complicated (see also ref.~\cite{SV}).

\noi
Our paper is organized as follows. In the next section we introduce the
chiral unitary matrix ensemble, and use the orthogonal polynomial method
to iteratively derive the relevant orthogonal polynomials for increasing
values of $N_f$. We prove that in the large-$N$ limit the $N$th orthogonal
polynomials 
$P_N^{(N_{f})}(\lambda;m_1^2,\ldots,m_{N_{f}}^2)$
for each $N_f$ have universal asymptotic limits near the
origin in both $\lambda = x/N^2$ 
and masses $m_f^2 = \mu_f^2/N^2$. 
{}From this it
follows that also all microscopic correlators, and the microscopic
spectral densities themselves, are universal. We also write down a simple
set of consistency relations for the microscopic spectral densities,
which follows from the decoupling of heavy fermions. These consistency
relations are found to be satisfied by our microscopic spectral 
densities. In section 3 we brief{}ly discuss some of the exact spectral sum 
rules for massive Dirac operators of QCD-like theories, and we verify that 
our double-microscopic spectral densities are consistent with these exact
sum rules, which have been derived without the use of random matrix theory.
As in the massless case \cite{V,VZ,us}, this provides strong support to the
conjecture that the microscopic spectral densities are universal not only
within the context of random matrix theory, but indeed are exact expressions
also for the full spectral densities of QCD in that limit. Section 4 
contains our conclusions, and a proof of the general-$N_f$ expressions
can be found in the appendix.

\section{The Chiral Unitary Ensemble}
Consider 4-dimensional $SU(N_c\!\geq\!3)$ gauge theories coupled to
$N_f$ fermions in the fundamental representation of the gauge group.
Assume that a non-vanishing chiral condensate $\Sigma \equiv \langle
\bar{\psi}\psi\rangle$ has been formed.
According to the conjectures of ref.~\cite{V} the microscopic spectral
density 
\beq
\rho_S(\zeta) \equiv \lim_{V_{4}\!\to\!\infty} \frac{1}{V_{4}\Sigma}
\rho(\frac{\zeta}{V_4\Sigma})
\eeq
of the Dirac operator can be computed {\em exactly} in an ensemble of complex 
block hermitian $(2N\! +\! |\nu |)\!\times\! (2N\! +\! |\nu |)$ matrices $M$:
\beq
M ~=~ \left( \begin{array}{cc}
              0 & W^{\dagger} \\
              W & 0
              \end{array}
      \right) ~.
\eeq
The partition function is defined by
\beq
{\cal Z} ~=~ \int\! dW \prod_{f=1}^{N_{f}}{\det}\left(M + i\,m_f\right)~ 
\exp\left[-\frac{N}{2} {\rm tr}\, V(M^2)\right] ~.
\label{cZ}
\eeq
Here $W$ is a rectangular complex matrix of size 
$N\times(N\! +\! |\nu|)$. The integer $\nu$ is identified with
topological charge, and the space-time volume $V_4$ is identified with
$2N\! +\! |\nu|$, the size of the matrix $M$. The 
integration measure in eq.~(\ref{cZ}) is the Haar measure of $W$. 
{}From now on consider the
sector of zero topological charge $\nu$, which is the case most relevant for
comparison with Monte Carlo data of lattice gauge theory. However, the
general-$\nu$ case can be extracted from the formulas we shall give below
simply by setting $|\nu|$ 
fermion masses to zero.

\noi
Rewriting the matrix integral (\ref{cZ}) in terms of
the (positive definite) eigenvalues $\lambda_i$ of the hermitian
matrix $W^{\dagger}W$, one gets, after discarding an irrelevant
overall factor from the angular integrations \cite{TRM}:
\beq
{\cal Z} ~=~ \int_0^{\infty}\! \prod_{i=1}^N \left(d\lambda_i
\prod_{f}(\lambda_i + m_f^2)~
{\rm e}^{-NV(\lambda_i)}\right)\left|{\det}_{ij}
\lambda_j^{i-1}\right|^2 ~.
\eeq

\noi
In terms of the standard orthogonal polynomial method, we thus seek
polynomials 
$P_n^{(N_{f})}(\la;m_1^2,\ldots,m_{N_{f}}^2)$ 
orthogonal with respect to the weight functions
\beq
w(\lambda) = \prod_{f=1}^{N_{f}}(\lambda + m_f^2)~
{\rm e}^{-NV(\la)} ~.
\eeq
So far we have not specified the potential $V(\lambda)$, which can
be parametrized in a quite general way by 
\beq
V(\lambda) ~=~ \sum_{k\geq 1} \frac{g_{k}}{k}\lambda^k ~.
\eeq 
As was shown in ref.~\cite{us}, when all $m_f\!=\! 0$ the orthogonal 
polynomials have, for fixed $x \!=\! N^2\la$ and $t\!=\! n/N$, a universal 
limiting behavior. When the polynomials can be normalized according
to $P_n^{(N_f)}(0) \!=\! 1$ the limit is 
\beq
\lim_{N\to\infty} \left.P_n^{(N_f)}(\frac{x}{N^2})\right|_{n=Nt}
~=~ N_f!\frac{J_{N_{f}}(u(t)\sqrt{x})}{(u(t)\sqrt{x}/2)^{N_{f}}} ~,
\label{prevuni}
\eeq
with 
\beq
u(t) ~=~ \int_0^t\frac{dt'}{\sqrt{r(t')}} ~~~~~~~~~~{\mbox{\rm and}}~~~~~
t ~=~ \sum_k \frac{g_{k}}{2}\left(2k \atop k \right) r(t)^k ~. 
\eeq
The restriction to normalizability according to  $P_n^{(N_f)}(0) \!=\! 1$
for $n = Nt$ can be rephrased \cite{us} as $\rho(0) \!\neq\! 0$, 
so the condition is
precisely as expected: the macroscopic spectral density must be 
non-vanishing at the origin. One can then identify
\beq 
a ~=~ 2\sqrt{r(1)} ~,~~~~~~~~~~\rho(0) ~=~ \frac{u(1)}{2\pi} ~,\label{urrel}
\eeq
where $a$ gives the upper limit of the support of $\rho(\la)$. Once
universality of the orthogonal polynomials has been established, the 
general proof of universality of all microscopic spectral correlators
follows as a simple corollary \cite{us}. 

\subsection{\sc One massive flavor}

To generalize these results to the case of massive fermions, we first note
that for quenched fermions $N_f\!=\!0$ there is no distinction between the
massive and massless cases. Consider next the first non-trivial case
of $N_f\!=\! 1$. We here need polynomials orthogonalized according to
\beq
\int_0^{\infty}d\la\, (\la+m^2)~{\rm e}^{-NV(\la)}P_i^{(1)}(\la;m^2)
P_j^{(1)}(\la;m^2) ~=~ h_i^{(1)} \delta_{ij} ~.\label{pnorm}
\eeq
Our basic observation is that these polynomials are readily found by
expanding in terms of the polynomials of the quenched case, 
$P_j^{(0)}(\la)$ by use of Christoffel's theorem \cite{SZE}:
\beq
P_i^{(1)}(\la;m^2) ~=~ \sum_{j=0}^i 
\frac{P_j^{(0)}(-m^2)}{h_j^{(0)}} P_j^{(0)}(\la) ~,
\eeq
up to an arbitrary normalization condition. By the Christoffel-Darboux 
formula we can thus choose
\beq
P_i^{(1)}(\la;m) ~=~ \frac{P_i^{(0)}(-m^2)P_{i+1}^{(0)}(\la)
- P_{i+1}^{(0)}(-m^2)P_i^{(0)}(\la)}{\la + m^2} ~,\label{Darboux1}
\eeq
and fix the normalization subsequently. Note that the r.h.s., despite
appearances, indeed is a polynomial in $\la$.

\noi
Because the new factor 
$(\la+m^2)$ in the weight function $w(\la)$ can be viewed as a contribution 
to the generic potential $V(\la)$ which is subdominant in $1/N$, the 
macroscopic spectral density is unchanged. Since $\rho(0)\neq 0$ this means 
that $P_N^{(1)}(0;m^2) \neq 0$ as well in the large-$N$ limit, and by 
redefining, for sufficiently large index $i$,
\beq
P_i^{(1)}(\la;m^2) ~\to~ P_i^{(1)}(\la;m^2)/P_i^{(1)}(0;m^2) ~,
\eeq
the orthogonal polynomials of eq.~(\ref{Darboux1}) become normalized 
according to $P_i^{(1)}(0;m^2)\!=\! 1$. The normalization constants 
$h_i^{(1)}$ of course become modified accordingly. 

\noi
In the double-microscopic limit where $x \!=\! N^2\la$ and $\mu \!=\! Nm$ 
are kept fixed as $N \!\to\! \infty$ we thus get from eq.~(\ref{Darboux1})
and the previous universal result (\ref{prevuni}):
\beq
\lim_{N\to\infty}
\left.
P_n^{(1)}(\frac{x}{N^2};\frac{\mu^2}{N^2})
\right|_{n=Nt}
\ ~=~
\frac{\mu^2}{x+\mu^2}\,
\left( 
J_0(u(t)\sqrt{x}) +
\frac{\sqrt{x}\, J_1(u(t)\sqrt{x})\,I_0(u(t)\mu)}{\mu\,{I_1(u(t)\mu)}}\right)
\eeq
where $J_n(x)$ and $I_n(x)$ are ordinary and modified Bessel functions,
respectively.
This depends on the given potential $V(\la)$ only implicitly through
the two functions $u(t)$ and $r(t)$, which are inherited from the massless 
case. Because of relation (\ref{urrel}), we have therefore proved that
the double-microscopic limit of the polynomial $P_N^{(1)}(x/N^2;\mu^2/N^2)$
is {\em universal}, depending only on the end-point of
the cut $a$ (which one can choose at some fixed value), and the value
of the macroscopic spectral density at the origin, $\rho(0)$. In applications
to Dirac spectra, this value is, by the Banks-Casher relation
\cite{BC}, fixed by
the chiral condensate: 
$\rho(0) \!=\! \langle\bar{\psi}\psi\rangle/\pi$.
As a first simple check we note that $P_n^{(1)}(x/N^2,\mu^2/N^2) \to
J_1(u(t)\sqrt{x})/(u(t)\sqrt{x})$ as $\mu\!\to\!0$, and we hence
recover the result of ref.~\cite{us}. In the opposite limit, $\mu\!\to\!
\infty$, we see that the above universal polynomial approaches
$J_0(u(t)\sqrt{x})$, the result of the quenched case. This is the first
example of the decoupling of a massive fermion, which we shall return to
in greater generality below.
 
\noi
It now follows as a simple corollary that also all microscopic spectral
correlators are universal in the above scaling limit. Consider first the 
kernel
\begin{eqnarray}
K^{(1)}_N(z_1,z_2;m) &=& \sqrt{|z_1\,z_2|}\,
\sqrt{{\rm e}^{-N(V(z_1^2)+V(z_2^2))} \, (z_1^2+m^2)(z_2^2+m^2)}
\nonumber\\
&\times& \frac{
P^{(1)}_{N-1}(z_1^2;m^2)P^{(1)}_{N}(z_2^2;m^2)-
P^{(1)}_{N}(z_1^2;m^2)P^{(1)}_{N-1}(z_2^2;m^2)}{z_1^2-z_2^2}
\label{CKone}
\end{eqnarray}
governing the correlation of $M$.
In the scaling limit it becomes of the universal form
\begin{eqnarray}
K_S^{(1)}(\zeta,\,\zeta'; \mu) &=&
\lim_{N\rightarrow\infty}
\frac{1}{N}\,{K}^{(1)}_N(\frac{\zeta}{N},\frac{\zeta'}{N};
\frac{\mu}{N}) \cr
&=&
C(\mu)\,
\sqrt{|\zeta\,\zeta'|}\,
\frac{
\det
\left(
\begin{array}{lll}
J_0(2\pi\rho(0)\zeta) &\zeta\, J_1(2\pi\rho(0)\zeta) & 
\zeta^2\,J_2(2\pi\rho(0)\zeta)\\
J_0(2\pi\rho(0)\zeta')&\zeta'\, J_1(2\pi\rho(0)\zeta') & {\zeta'}^2\, 
J_2(2\pi\rho(0)\zeta')\\
I_0(2\pi\rho(0)\mu)   &-\mu\, I_1(2\pi\rho(0)\mu) & \mu^2\,I_2(2\pi\rho(0)\mu)
\end{array}
\right)
}{
(\zeta^2-{\zeta'}^2) \sqrt{(\zeta^2+\mu^2) ({\zeta'}^2+\mu^2)}
} ~.
\end{eqnarray}
$C(\mu)$ is a function of $\mu$ only, to be determined below.

\noi
All double-microscopic spectral correlators
\beq
\rho^{(1)}_N(z_1,\ldots,z_s;m)=
\left\langle\prod_{a=1}^s \frac{1}{2N}\,{\rm tr}\,\delta(z_a-M) \right\rangle
=
\det_{a,b} K^{(1)}_N(z_a,z_b;m)
\eeq
therefore reach universal limits as well:
\beq
\rho^{(1)}_S(\zeta_1,\ldots,\zeta_s;\mu) = 
\lim_{N\to\infty}
\frac{1}{N^s}
\rho^{(1)}_N(\frac{\zeta_1}{N},\ldots,\frac{\zeta_s}{N};\frac{\mu}{N})=
\det_{a,b} K^{(1)}_S(\zeta_a,\zeta_a;\mu) ~.
\label{micspcorr}
\eeq
\noi
In particular, the spectral density itself,
\beq
\rho^{(1)}_N(z;m) ~=~ K^{(1)}_N(z,z;m)
\eeq
takes a universal form in the double-microscopic limit, which, after choosing
the conventional normalization \cite{V}, reads
\beq
\rho^{(1)}_S(\zeta;\mu) = 
\frac{1}{\pi\rho(0)}
K_S^{(1)}(
\frac{\zeta}{2\pi\rho(0)},
\frac{\zeta}{2\pi\rho(0)};\frac{\mu}{2\pi\rho(0)}) ~.
\label{mrhodef}
\eeq
After making use of some Bessel function identities, this can, 
be simplified to\footnote{Our expression for the
double-microscopic spectral density does not agree with that of ref.
\cite{JNZ}, where a derivation was given in terms of a saddle-point
evaluation of a matrix model with Gaussian action. Because of this 
disagreement, we have taken special pains to check that the result presented
here is correct. While both the result presented in ref.~\cite{JNZ} and
the expression shown here (eq.~(\ref{r1})) reduce to the previously known
result in the massless limit, only the expression (\ref{r1}) reduces
correctly to the quenched result in the limit $\mu \to \infty$ (see below).
One further check on the result (\ref{r1}) comes from the fact that it
correctly reproduces an exact spectral sum rule of QCD; see next section.}
\beq
\rho_S^{(1)}(\zeta;\mu) = \frac{|\zeta|}{2}
\left(J_0(\zeta)^2 + J_1(\zeta)^2\right)
-|\zeta|\frac{
J_0(\zeta)[\mu\,I_1(\mu)\,J_0(\zeta)+I_0(\mu)\,\zeta\,J_1(\zeta)]}
{(\zeta^2+\mu^2)\,I_0(\mu)}~.
\label{r1}
\eeq

\noi
The microscopic spectral correlators, and the density itself, 
have had their overall normalization fixed by the
matching condition (the compensating factor of $\pi\rho(0)$ on the left
hand side of the relation below is due to the normalization convention 
(\ref{mrhodef})) between micro- and macroscopic densities, 
\beq
\lim_{\zeta\to\infty}[\pi\rho(0)\,\rho_S^{(1)}(\zeta;\mu)] ~=~ \rho(0) ~,
\eeq
the latter of which is
independent of $\mu$. This is just a convenient short-cut. We could
have avoided it by explicitly computing the normalization constants
$h_n^{(1)}$ of eq.~(\ref{pnorm}), but which we see no need to do that here.
In addition we of course have that $\rho_S^{(1)}(\zeta;0)$ 
agrees with the result obtained directly from the massless case \cite{V}. 
A more interesting condition comes from the decoupling of a very massive
fermion. This implies
\beq
\lim_{\mu\to\infty}\rho_S^{(1)}(\zeta;\mu) ~=~ \rho_S^{(0)}(\zeta) ~.
\label{decoup1}
\eeq
One verifies that indeed this relation is satisfied by the density 
(\ref{r1}). 

\subsection{\sc More massive flavors}

Next, consider the case of more quark flavors. The decoupling of heavy
fermions, of which we already saw one example in eq.~(\ref{decoup1}), leads
to a hierarchy of consistency relations which must be satisfied by the
microscopic spectral densities:
\begin{eqnarray}
\rho_S^{(N_{f})}(\zeta;\mu_1,\ldots,\mu_{N_{f}}) &\to&
\rho_S^{(N_{f}-1)}(\zeta;\mu_1,\ldots,\mu_{N_{f}-1}) ~~~~~~{\mbox{\rm as}}
~~~ \mu_{N_{f}} \to \infty \cr
\rho_S^{(N_{f})}(\zeta;\mu_1,\ldots,\mu_{N_{f}}) &\to&
\rho_S^{(N_{f}-2)}(\zeta;\mu_1,\ldots,\mu_{N_{f}-2}) ~~~~~~{\mbox{\rm as}}
~~~ \mu_{N_{f}-1},\mu_{N_{f}} \to \infty \cr
\vdots ~~~~~~~~~~~~~ && ~~~~ \vdots \cr
\rho_S^{(N_{f})}(\zeta;\mu_1,\ldots,\mu_{N_{f}}) &\to&
\rho_S^{(0)}(\zeta) ~~~~~~~~~~~~~~~~~~~~~~~~~~~~~~~{\mbox{\rm as}}
~~~ \mu_1,\ldots,\mu_{N_{f}} \to \infty ~,\label{gendecoup}
\end{eqnarray}
and all other relations obtained by permutations of the $\mu_i$. There are 
similar consistency relations for all microscopic spectral correlators.
In addition, we of course also have that in the limit where all masses
are set to zero, the result must agree with that of ref.~\cite{V}. 

\noi
We construct higher-$N_f$ microscopic spectral densities by straightforward
iteration of the procedure described above. For $N_f\!=\!2$ the new
orthogonal polynomials become
\beq
P_i^{(2)}(\la;m_1^2,m_2^2) ~=~ 
\frac{P_i^{(1)}(-m_2^2;m_1^2)P_{i+1}^{(1)}(\la;m_1^2)
- P_{i+1}^{(1)}(-m_2^2;m_1^2)P_i^{(1)}(\la;m_1^2)}{\la + m_2^2} 
~,\label{Darboux2}
\eeq
which we again can give our conventional normalization by letting
\beq
P_i^{(2)}(\la;m_1^2,m_2^2) ~\to~ 
P_i^{(2)}(\la;m_1^2,m_2^2)/P_i^{(2)}(0;m_1^2,m_2^2) ~.
\eeq
Taking the large-$N$ limit with $x \!=\! N^2\la ,~\mu_1 \!=\! Nm_1$ and  
$\mu_2 \!=\! Nm_2$ fixed, leads to
\beq
\lim_{N\to\infty}
\left.
P_n^{(2)}(\frac{x}{N^2};\frac{\mu_1^2}{N^2},\frac{\mu_2^2}{N^2}) 
\right|_{n=Nt}
~=~
\frac{\mu_1^2}{x+\mu_1^2}\frac{\mu_2^2}{x+\mu_2^2}
\frac{
\det
\left(
\begin{array}{lll}
J_0(u(t)\sqrt{x})  &\sqrt{x}\,J_1(u(t)\sqrt{x}) & x\, J_2\,(u(t)\sqrt{x})\\
I_0(u(t)\mu_1)  &-\mu_1\,I_1(u(t)\mu_1) & \mu_1^2 I_2\,(u(t)\mu_1)\\
I_0(u(t)\mu_2)  &-\mu_2\,I_1(u(t)\mu_2) & \mu_2^2 I_2\,(u(t)\mu_2)
\end{array}
\right)
}{
\det
\left(
\begin{array}{ll}
-\mu_1\,I_1(u(t)\mu_1) &  \mu_1^2 I_2\,(u(t)\mu_1)\\
-\mu_2\,I_1(u(t)\mu_2) &  \mu_2^2 I_2\,(u(t)\mu_2)
\end{array}
\right)
} .
\eeq
Also here, and for the appropriate generalization to all higher values of 
$N_f$, the universality proof of ref.~\cite{us} immediately implies
that $P_n^{(2)}(x/N^2;\mu_1^2/N^2,\mu_2^2/N^2)$ is universal. All 
double-microscopic
spectral correlators (\ref{micspcorr}) therefore are universal as well.
The results (generalized Bessel kernels) can immediately be written down, 
following the definitions (\ref{micspcorr}). Most compact expressions
seem to be obtained by writing everything exclusively in terms of the
zeroth and first (modified) Bessel functions. We restrict ourselves to
displaying only the most important of them, the double-microscopic
spectral density itself, which becomes
\begin{eqnarray}
\rho_S^{(2)}(\zeta;\mu_1,\mu_2)\! &=& \frac{|\zeta|}{2}\!\left(
J_0(\zeta)^2 + J_1(\zeta)^2\right)  - 
\label{r2}\\
&& \frac{|\zeta|(\mu_1^2 \!-\! \mu_2^2)}{(\zeta^2\!+\!\mu_1^2)
(\zeta^2\!+\!\mu_2^2)}
\frac{[\mu_1 I_1(\mu_1)J_0(\zeta) \!+\! \zeta I_0(\mu_1)J_1(\zeta)] 
[\mu_2 I_1(\mu_2)J_0(\zeta) \!+\! \zeta I_0(\mu_2)J_1(\zeta)]}
{\mu_1 I_1(\mu_1)I_0(\mu_2) - \mu_2 I_0(\mu_1)I_1(\mu_2)} ~.
\nonumber
\end{eqnarray}
We note that it satisfies all the required consistency conditions 
(\ref{gendecoup}).

\noi
The iterative procedure described above gives the relevant orthogonal 
polynomials, and hence all the microscopic spectral correlators
and densities for arbitrary $N_f$.
For example, the orthogonal polynomial for $N_f\!=\!3$ reads
\bea
&&\lim_{N\to\infty}
\left. 
P_n^{(3)}(\frac{x}{N^2};
\frac{\mu_1^2}{N^2},\frac{\mu_2^2}{N^2},\frac{\mu_3^2}{N^2})
\right|_{n=Nt}=\\
&&
\frac{\mu_1^2}{x+\mu_1^2}\frac{\mu_2^2}{x+\mu_2^2}
\frac{\mu_3^2}{x+\mu_3^2}
\frac{\det\left(
\begin{array}{llll}
J_0(u(t)\sqrt{x})  &\sqrt{x}\,J_1(u(t)\sqrt{x}) & x\, J_2(u(t)\sqrt{x})\, 
& x^{3/2}J_3(u(t)\sqrt{x})\\
I_0(u(t)\mu_1)  &-\mu_1\,I_1(u(t)\mu_1) & \mu_1^2 I_2(u(t)\mu_1) & -
\mu_1^3 I_3(u(t)\mu_1)\\
I_0(u(t)\mu_2)  &-\mu_2\,I_1(u(t)\mu_2) & \mu_2^2 I_2(u(t)\mu_2) & -
\mu_2^3 I_3(u(t)\mu_2)\\
I_0(u(t)\mu_3)  &-\mu_3\,I_1(u(t)\mu_3) & \mu_3^2 I_2(u(t)\mu_3) & -
\mu_3^3 I_3(u(t)\mu_3)
\end{array}
\right)
}{
\det
\left(
\begin{array}{lll}
-\mu_1\,I_1(u(t)\mu_1) & \mu_1^2 I_2(u(t)\mu_1) & -\mu_1^3 I_3\,(u(t)\mu_1)\\
-\mu_2\,I_1(u(t)\mu_2) & \mu_2^2 I_2(u(t)\mu_2) & -\mu_2^3 I_3\,(u(t)\mu_2)\\
-\mu_3\,I_1(u(t)\mu_3) & \mu_3^2 I_2(u(t)\mu_3) & -\mu_3^3 I_3\,(u(t)\mu_3)
\end{array}
\right)
}~.
\nonumber
\eea
We shall give the general expression for spectral correlators
for an arbitrary number of flavors $N_f$ below. In fig.~1 we show the
microscopic spectral density $\rho_S^{(3)}(\zeta;\mu,\mu,\mu)$
with, for simplicity, degenerate masses. The decoupling
of heavy fermions that is expressed by eq.~(\ref{gendecoup}) is easily seen 
in the convergence towards the quenched result $\rho_S^{(0)}(\zeta)$,
while in the opposite limit of zero masses the density approaches the
result of ref.~\cite{V}.

\centerline{\epsfbox{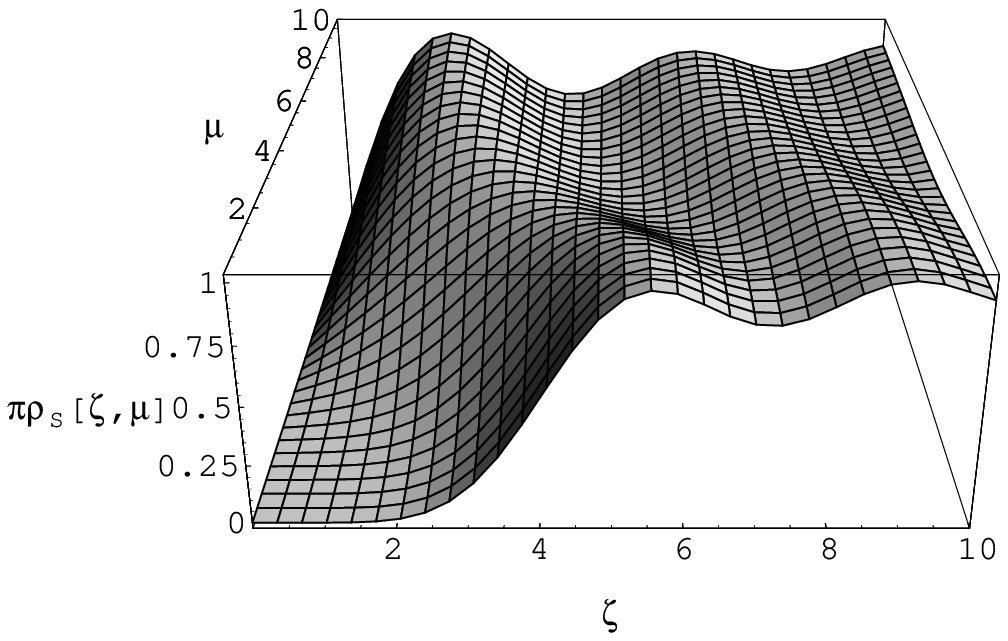}}
\centerline{{\small Figure 1: Microscopic spectral density for 3 flavors
with degenerate masses.}}

\noi
Before proceeding to the general formula for arbitrary $N_f$, we note that
there are still a few particular cases that can be expressed in a very 
simple analytical form. As an example, consider the case of {\em one} massive
fermion (of rescaled mass $\mu$), and $N_f\!-\!1$ massless ones. 
For $N_f\!=\!1$
this case of course coincides with what we have already described above.
For general $N_f$ the formula can worked out completely analogous to the
case of one massive flavor,
by starting from the asymptotic form of the polynomial (\ref{prevuni}). 
The resulting double-microscopic spectral density
reads as follows:
\beq
\rho_S^{(N_{f})}(\zeta;\mu,0,\cdots,0) = \frac{|\zeta|}{2}\!\left\{\!
J_{N_{f}}(\zeta)^2 \!-\! J_{N_{f}-1}(\zeta)J_{N_{f}+1}(\zeta)
+ \frac{\mu^2 J_{N_{f}-1}(\zeta)}{N_{f}(\zeta^2 + \mu^2)}\!\left(\!
\frac{I_{N_{f}+1}(\mu)}{I_{N_{f}-1}(\mu)}J_{N_{f}-1}(\zeta)\!
+\! J_{N_{f}+1}(\zeta)\right)\!\right\} ~.\label{rmassless}
\eeq
One verifies that also this microscopic spectral densities satisfies the
consistency condition (\ref{gendecoup}).

\subsection{\sc An arbitrary number of massive flavors}

For practical purposes, only the massive microscopic spectral densities
for a few flavors may be of relevance. Nevertheless, it is worthwhile to note
that the iterative procedure described above can be carried through for
an arbitrary number of flavors $N_f$, which we here for notational
convenience denote by $\alpha$. 

\noi 
We shall first state the main results. The orthogonal polynomials, 
normalized by 
$$
P^{(\alpha)}_n(0;\frac{\mu_1^2}{N^2},\cdots,
\frac{\mu_{\alpha}^2}{N^2})\!=\! 1 ~,
$$
read
\bea
&&\lim_{N\to\infty}
\left.
P^{(\alpha)}_n(\frac{x}{N^2};\frac{\mu_1^2}{N^2},\cdots,
\frac{\mu_{\alpha}^2}{N^2})
\right|_{n=Nt}=
\nonumber\\
&&
\prod_{f=1}^{\alpha}\frac{\mu_f^2}{x+\mu_f^2} \,
\frac{
\det
\left(
\begin{array}{cccc}
J_0(u(t)\sqrt{x})  &\sqrt{x}\, J_1(u(t)\sqrt{x}) & \cdots 
& x^{(\alpha+1)/2}\, J_{\alpha+1}(u(t)\sqrt{x}) \\
I_0(u(t)\mu_1)  &-\mu_1\, I_1(u(t)\mu_1) & \cdots 
&(-\mu_1)^{\alpha+1}\, I_{\alpha+1}(u(t)\mu_1) \\
\vdots    & \vdots   & \cdots           & \vdots \\
I_0(u(t)\mu_\alpha)  &-\mu_\alpha\, I_1(u(t)\mu_\alpha) & \cdots 
&(-\mu_1)^{\alpha+1}\, I_{\alpha+1}(u(t)\mu_\alpha) 
\end{array}
\right)}{
\det_{i,j}
(-\mu_i)^{j} I_{j}(u(t)\mu_i) } ~.
\label{genpol}
\eea
The double-microscopic spectral correlators
\beq
\rho^{(\alpha)}_{S}(\zeta_1,\cdots,\zeta_s;\mu_1,\cdots,\mu_{\alpha})\equiv
\lim_{N\rightarrow\infty}\frac{1}{N^s}\,
\rho^{(\alpha)}_N(\frac{\zeta_1}{N},\cdots,\frac{\zeta_s}{N};
\frac{\mu_1}{N},\cdots,\frac{\mu_{\alpha}}{N})
=
\det_{a,b} K^{(\alpha)} (\zeta_a,\zeta_b;\{\mu_f\}),
\eeq
are expressed through the kernel which is given by
\bea
&&K^{(\alpha)}(\zeta,\zeta';\{\mu_f\})=
\label{generalK}\\
&&-\pi\rho(0)
\frac{\sqrt{|\zeta\,\zeta'|}}{\zeta^2-{\zeta'}^2}
\frac{
\det
\left(
\begin{array}{cccc}
J_0(2\pi\rho(0)\zeta)  &\zeta\, J_1(2\pi\rho(0)\zeta) & \cdots 
& \zeta^{\alpha+1}\, J_{\alpha+1}(2\pi\rho(0)\zeta) \\
J_0(2\pi\rho(0)\zeta')  & \zeta'\, J_1(2\pi\rho(0)\zeta') & \cdots & 
{\zeta'}^{\alpha+1}\,J_{\alpha+1}(2\pi\rho(0)\zeta') \\
I_0(2\pi\rho(0)\mu_1)  &-\mu_1\, I_1(2\pi\rho(0)\mu_1) & \cdots 
&(-\mu_1)^{\alpha+1}\, I_{\alpha+1}(2\pi\rho(0)\mu_1) \\
\vdots    & \vdots   & \cdots           & \vdots \\
I_0(2\pi\rho(0)\mu_\alpha)  &-\mu_\alpha\, I_1(2\pi\rho(0)\mu_\alpha) & \cdots 
&(-\mu_{\alpha})^{\alpha+1}\, I_{\alpha+1}(2\pi\rho(0)\mu_\alpha) 
\end{array}
\right)
}{
\prod_{f = 1}^{\alpha}
\sqrt{
\left( {{\zeta }^2} + {{{{\mu }_f}}^2} \right) \,
( {{\zeta' }^2} + {{{{\mu }_f}}^2} ) 
}\,
\det_{i,j}
(-\mu_i)^{j-1} I_{j-1}(2\pi\rho(0)\mu_i) }~.
\nonumber
\eea
The double-microscopic spectral density thus explicitly reads
\beq
\rho^{(\alpha)}_S (\zeta ; \{\mu_f\})=
\frac{-|\zeta|}{2}
\frac{\det
\left(
\begin{array}{cccc}
\zeta^{-1}J_{-1}(\zeta)  & J_0(\zeta) & \cdots 
& \zeta^{\alpha}\, J_{\alpha}(\zeta) \\
J_0(\zeta)  & \zeta\, J_1(\zeta) & \cdots  
& {\zeta}^{\alpha+1}\,J_{\alpha+1}(\zeta) \\
I_0(\mu_1)  &-\mu_1\, I_1(\mu_1) & \cdots 
&(-\mu_1)^{\alpha+1}\, I_{\alpha+1}(\mu_1) \\
\vdots    & \vdots   & \cdots           & \vdots \\
I_0(\mu_\alpha)  &-\mu_\alpha\, I_1(\mu_\alpha) & \cdots 
&(-\mu_{\alpha})^{\alpha+1}\, I_{\alpha+1}(\mu_\alpha) 
\end{array}
\right)
}{\prod_{f = 1}^{\alpha}
\left( {{\zeta }^2} + {{{{\mu }_f}}^2} \right)\, 
\det {\cal M}} ~.
\label{rho}
\eeq
Here ${\cal M}$ is an $\alpha\!\times\!\alpha$ matrix defined by
${\cal M}_{ij} \!=\! (-\mu_i)^{j-1}I_{j-1}(\mu_i)$.
The proof of the above formulae is summarized in Appendix A.

\noi
We can now easily check that
the microscopic kernel (\ref{generalK}) and density (\ref{rho})
for arbitrary $\alpha$ satisfies all the
decoupling relations (\ref{gendecoup}).
Suppose $\mu_\alpha$ is taken to be infinity in eq. (\ref{rho}).
Due to the asymptotic behavior of the modified Bessel functions
\beq
I_i(\mu)~\sim~ \frac{{\rm e}^\mu}{\sqrt{2\pi\mu}}~~~~~~~~~
{\mbox{\rm as}}~~~\mu\to\infty ~,
\eeq
the determinant in the numerator of (\ref{rho}) is dominated by
its minor for the lower right corner,
\beq
(-\mu_\a)^{\a+1}\,
\frac{{\rm e}^{\mu_\a}}{\sqrt{2\pi\mu_\a}}
\,\det
\left(
\begin{array}{cccc}
\zeta^{-1}J_{-1}(\zeta) & J_0(\zeta) & \cdots & \zeta^{\a-1}J_{\alpha-1} (\zeta) \\
J_0(\zeta) & \zeta\,J_{1}(\zeta) & \cdots & \zeta^{\a}J_{\alpha} (\zeta) \\
I_0(\mu_1) & -\mu_1\, I_{1}(\mu_1) & \cdots & (-\mu_1)^{\a}I_{\alpha} (\mu_1) \\
\vdots    & \vdots   & \cdots & \vdots \\
I_0(\mu_{\a-1}) & -\mu_{\a-1}\, I_{1}(\mu_{\a-1}) & \cdots &
 (-\mu_{\a-1})^{\a}I_{\alpha} (\mu_{\a-1}) 
\end{array}
\right) ~.
\eeq
Similarly, its denominator is approximated by
\beq
\mu_\a^2\,\prod_{f = 1}^{\alpha-1}
\left( {{\zeta }^2} + {{{{\mu }_f}}^2} \right) \cdot
(-\mu_a)^{\a-1}
\frac{{\rm e}^{\mu_\a}}{\sqrt{2\pi\mu_\a}}
\,
\det_{1\leq i,j\leq \alpha-1}\,
(-\mu_i)^{j-1} {I_{j-1}}({{\mu }_i}) ~.
\eeq
Thus we recover the same expression (\ref{rho}) for 
$\alpha\to\alpha-1$.
By iteration, we confirm that the tower of consistency relations
(\ref{gendecoup}) is satisfied.

\noi
On the other hand, we have another consistency relation
that when all masses vanishes, $\rho_S^{(\alpha)}(\zeta;0,\cdots,0)$ 
should agree with the result obtained directly from the massless case \cite{V}. 
In order to check this, let us take the limit where 
the masses are taken to be zero one by one.
Suppose $\mu_\alpha$ is taken to be zero first.
Since
\beq
I_0(0)=1~,~~~~~ I_n(0)=0\ \ (n\geq 1)~,
\eeq
the determinants
in the numerator and denominator of (\ref{rho}) are replaced by
their minors for the lower left corner.
Then we obtain
\bea
&&\rho^{(\alpha)}_S (\zeta ; \mu_1, \cdots,\mu_{\alpha-1},0)=\nonumber\\
&&\frac{-|\zeta|}{2}
\frac{\det
\left(
\begin{array}{cccc}
J_0(\zeta) & \cdots & \zeta^{\a-1}J_{\alpha-1} (\zeta) &
 \zeta^{\a}J_{\alpha} (\zeta)\\
J_1(\zeta) & \cdots & \zeta^{\a}J_{\alpha} (\zeta) & 
\zeta^{\a+1}J_{\alpha+1} (\zeta)\\
-\mu_1\, I_{1}(\mu_1) & \cdots & (-\mu_1)^{\a}I_{\alpha} (\mu_1) & 
(-\mu_1)^{\a+1}I_{\alpha+1} (\mu_1) \\
\vdots & \cdots & \vdots & \vdots \\
-\mu_{\a-1}\, I_{1}(\mu_{\a-1}) & \cdots & 
(-\mu_{\a-1})^{\a}I_{\alpha} (\mu_{\a-1}) & 
(-\mu_{\a-1})^{\a+1}I_{\alpha+1} (\mu_{\a-1}) \\
\end{array}
\right)
}{{\zeta }^2\,\prod_{f = 1}^{\alpha-1}
\left( {{\zeta }^2} + {{{{\mu }_f}}^2} \right)\,
\det_{1\leq i,j\leq \alpha-1}\,
(-\mu_i)^{j} {I_{j}}({{\mu }_i})} ~.
\eea
We may iterate this procedure as many times as required
to obtain,
\beq
\rho^{(\alpha)}_S (\zeta ; \mu, 0,\cdots,0)=
\frac{-|\zeta|}{2}
\frac{\det
\left(
\begin{array}{lll}
\zeta^{\a-2} J_{\alpha-2}(\zeta) &\zeta^{\a-1}  J_{\alpha-1}(\zeta) & \zeta^{\a} J_{\alpha}(\zeta) \\
\zeta^{\a-1} J_{\alpha-1}(\zeta) &\zeta^{\a}  J_{\alpha}(\zeta) & \zeta^{\a+1} J_{\alpha+1}(\zeta) \\
(-\mu)^{\a-1} I_{\alpha-1}(\mu) & (-\mu)^{\a} I_{\alpha}(\mu)   & (-\mu)^{\a+1} I_{\alpha+1}(\mu)
\end{array}
\right)
}{{\zeta }^{2(\alpha-1)}\,
\left( {{\zeta }^2} + {{{{\mu }}}^2} \right)
\, 
(-\mu)^{\a-1} {I_{\alpha-1}}({{\mu }})} ~,
\label{1massiverho}
\eeq
\bea
\rho^{(\alpha)}_S (\zeta ; 0,\cdots,0) &=&
\frac{-|\zeta|}{2}
\frac{\det
\left(
\begin{array}{ll}
\zeta^{\a-1}J_{\alpha-1}(\zeta) & \zeta^{\a}J_{\alpha}  (\zeta)\\
\zeta^{\a}J_{\alpha}(\zeta)   & \zeta^{\a+1}J_{\alpha+1}(\zeta)
\end{array}
\right)}{\zeta^{2\alpha}} \cr
&=& \frac{|\zeta|}{2} (J_\alpha(\zeta)^2-J_{\alpha-1}(\zeta)J_{\alpha+1}
(\zeta)) ~.
\label{masslessrho}
\eea
We note that the expression (\ref{1massiverho}) coincides with that of
eq. (\ref{rmassless}).
The last expression, (\ref{masslessrho}) indeed also agrees with the known 
formula for massless quarks \cite{V}.

\section{Exact Massive Spectral Sum Rules in QCD}

So far our discussion has been entirely within the framework of random
matrix theory. A highly non-trivial question is whether the resulting
universal spectral correlators are exact statements about the spectral
correlators in QCD as well. With the increasing evidence that this
is the case for massless Dirac operators, it is natural to expect that it
extends to massive Dirac operators as well, once considered in the appropriate
double-microscopic limit which we defined in the preceding section. We shall
here present some evidence that this is the case. We do this by temporarily
leaving the framework of random matrix models, and turning to the QCD
partition function, as it can be represented in the range $1/\Lambda_{{\rm QCD}}
\ll N^{1/d} \ll 1/m_{\pi}$. Here $N$ gives the size of the volume, and 
$\Lambda_{{\rm QCD}}$ is a typical hadronic scale in QCD. 
Following the conventional notation, we denote (up to a 
sign, see the discussion in ref.~\cite{LS}) the
chiral condensate by $\Sigma \equiv \langle \bar{\psi}\psi\rangle$.

\noi
Leutwyler and Smilga \cite{LS} have shown how to derive exact
spectral sum rules for massless Dirac operators in the above range.\footnote{
See also the work of ref.~\cite{SmV} for generalizations to gauge group 
$SU(2)$, and to more exotic flavor symmetry breaking patterns.} 
One generalization 
to the case of massive spectral sum rules has brief{}ly been considered by 
Shuryak and Verbaarschot \cite{SV}. Since the masses and the eigenvalues are
rescaled at the same rate in the $N \to \infty$ double-microscopic limit,
it is clear that we should consider spectral sum rules for which the
masses and the eigenvalues enter on equal footing. Consider first the case of
one flavor, $N_f\!=\!1$, in the sector of vanishing topological charge. 
The natural generalization of the spectral sum
rules in terms of inverse powers of $z^2$ is to consider inverse
powers of $z^2 + m^2$. From the expansion \cite{LS} of the QCD 
partition function in terms of Bessel functions in the above range, one 
finds that
\begin{eqnarray}
\frac{1}{\Sigma^2 N^2} \left\langle \sum_n~\!' \frac{1}{z_n^2 + m^2} 
\right\rangle &=& \sum_n \frac{1}{j_{0,n} + \mu^2} \cr
&=& \frac{I_1(\mu)}{2\mu I_0(\mu)} ~.
\end{eqnarray}
where we have defined $\mu \equiv mN\Sigma$. 
The sum on the right hand side runs
over the real zeros $j_{0,n}$ of the Bessel function $J_0(x)$. Defining
the double-microscopic spectral density 
\beq
\rho_S(\zeta;\mu) ~\equiv~ \lim_{N\to\infty} \frac{1}{\Sigma N}\rho\left(
\frac{\zeta}{\Sigma N}\right) ~, ~~~~~~~~ \mu = mN\Sigma ~~{\mbox{\rm   
fixed}}
\eeq
in terms of the ordinary macroscopic spectral density
\beq
\rho(z) ~=~ \langle \sum_n~\!' \delta(z - z_n)\rangle ~,
\eeq
one sees that the above spectral sum rule can be written
\beq
\int_0^{\infty} \! d\zeta~ \frac{\rho_S(\zeta;\mu)}{\zeta^2 + \mu^2} ~=~
\frac{I_1(\mu)}{2\mu I_0(\mu)} ~.
\eeq

\noi
This massive spectral sum rule, derived entirely within the framework of
QCD (a very simple chiral Lagrangian, in the above range), can now be
tested on the double-microscopic spectral density $\rho_S^{(1)}(\zeta;\mu)$
of eq.~(\ref{r1}). Performing the required integrals, one finds that it
indeed is satisfied. As in the massless case, the microscopic spectral
density of the chiral unitary ensemble is thus consistent with the spectral
sum rule of QCD.

\noi
Another example of an exact spectral sum rule which can be derived
directly from QCD, and for which we can compare with our present results, is
given by the case of one massive fermion of mass $m$, and $N_f\!-\! 1$
massless fermions. Again expanding the Bessel functions of the relevant
partition function of QCD with one massive flavor and $N_f\!-\!1$
massless flavors \cite{LS}, one finds the following exact spectral sum rule
($j_{m,n}$ denotes the $n$th zero of the Bessel function $J_m(x)$):
\begin{eqnarray}
\int_0^{\infty} \! d\zeta ~\frac{\rho_S(\zeta;\mu)}{\zeta^2 + \mu^2} &=&
\sum_n \frac{1}{j_{N_{f}-1,n} + \mu^2} \cr
&=& \frac{I_{N_{f}}(\mu)}{2\mu I_{N_{f}-1}(\mu)} ~.
\end{eqnarray}
Inserting the microscopic spectral density
$\rho_S^{(N_{f})}(\zeta;\mu,0,\ldots,0)$ of eq.~(\ref{rmassless}) we
verify that also this identity is satisfied. Other spectral sum rules
can be worked out analogously.

\section{Conclusions}
Motivated by the recent progress in numerically computing the microscopic
spectral densities of the Dirac operator in realistic four-dimensional
lattice gauge theories \cite{BBMSVW}, we have set out to compute the
microscopic spectral densities of what should correspond to massive Dirac
operators. Such an extension will presumably be required before a detailed
comparison with lattice gauge theory data can be performed beyond the
quenched approximation. To reach the universal limit, masses must be
scaled at a very precise rate as the volume is increased, but this
is entirely feasible in the context of lattice gauge theory. We thus
expect that the results presented here may be of practical value when
it comes to detailed comparisons between lattice Monte Carlo data for
QCD and the universality predictions from random matrix theory.

\noi
We have succeeded in deriving the relevant microscopic (we have called
these ``double-microscopic'' because both masses and eigenvalues are rescaled
as $N \to \infty$) spectral correlators, and in particular
the microscopic spectral densities themselves, within the framework of
random matrix models. In doing so, we have simultaneously extended the
universality proofs of ref.~\cite{us} to this more general 
situation.\footnote{An alternative route would be to derive directly the 
differential equation that the orthogonal polynomials satisfy in the 
microscopic limit, for example using the method of Kanzieper 
and Freilikher \cite{KF}, or to use the addition formalism of Zee \cite{Zee}.} 
The case of the unitary matrix ensemble, conjectured to be possibly
relevant for $SU(N_c)$ gauge theories in $(2\!+\!1)$ dimensions
\cite{VZ}, can be worked out analogously and will be presented
elsewhere \cite{DN}. As in ref.~\cite{us}, the proven universality is 
strictly limited to the framework of random matrix theory.  

\noi
Based on the decoupling of heavy fermions, we have derived a set of
consistency conditions for the microscopic spectral densities. These are
non-trivial relations that show how the various microscopic spectral
densities $\rho_S^{(N_{f})}(\zeta;\mu_1,\ldots,\mu_{N_{f}})$ for different
values of $N_f$ must be related to each other as one or more of the masses 
are sent to infinity. We have verified that our spectral densities
satisfy all of these general consistency conditions. In the other extreme
limit where all masses are taken to zero, we of course recover the known
results \cite{V,VZ}. 

\noi
To investigate the question as to whether QCD with massive flavors indeed
fall into the universality classes derived here, we have confirmed that
exact massive spectral sum rules (generalizations of the Leutwyler-Smilga
sum rules \cite{LS,SV}) derived directly from QCD are satisfied if we
identify the double-microscopic spectral density of QCD with that of
the chiral unitary ensemble. This, together with the mounting evidence
from the massless case, makes it highly plausible that all the universal 
double-microscopic spectral correlators we have derived here are exact
statements about QCD in the above limit. The crucial test will be a
comparison with results from lattice gauge theory, where these 
microscopic spectral correlators and densities are very convenient
finite-size scaling functions, almost tailored for Monte Carlo simulations.
Hopefully such data will soon be available. 
\vspace{0.5cm}

\noi
{\Large \bf{Acknowledgements}}
\nopagebreak

\noi
We thank M. Nowak and T. Wettig for stimulating discussions.
The work of S.M.N. is supported in part by the Nishina Memorial Foundation
and by NSF Grant PHY94-07194.

\noi
{\bf Note added:} Our main results (33) and (34) of this paper have been 
derived independently for the case of the Laguerre ensemble by Wilke, Guhr and
Wettig (hep-th/9711057), in a paper which appeared on the e-print
archives just a few days after ours (these authors did not consider the 
question of universality). 
The sum rule (45) has been derived earlier by Gasser and
Leutwyler in Phys. Lett. {\bf B188} (1987) 477, and has been compared
to lattice QCD data by Verbaarschot in Phys. Lett. {\bf B368} (1996) 137. 
The authors of ref. \cite{JNZ} have found an error in their previous
calculation of the $N_f=1$ case in the Gaussian ensemble, and they now
agree with the expression given in eq. (23) 
(M. Nowak, private communication).

\vspace{0.5cm}

\setcounter{equation}{0}
\renewcommand{\theequation}{A.\arabic{equation}}
\renewcommand{\thesection}{A}
\section{Proof of the Formulae in section 2.3}
Since the procedure of constructing the kernel
out of the polynomials (\ref{CKone})
is essentially the same as that of
adding another massive flavor (\ref{Darboux1})
up to the sign of $m^2$, it suffices to prove (\ref{generalK}).
We first demonstrate the following lemma:

\noi
{\bf Lemma} :\\
Let $P^{[\alpha]}(t;\lambda_0,\lambda_1,\cdots,\lambda_\alpha)$,
$\alpha=0,1,2,\cdots$, be
a set of functions generated by the iteration
\bea
&&P^{[\alpha+1]}(t;\lambda_0,\lambda_1,\cdots,\lambda_{\alpha+1})=\\
&&\frac{
P^{[\alpha]}(t;\lambda_0,\lambda_1,\cdots,\lambda_{\alpha})
P_t^{[\alpha]}(t;\lambda_{\alpha+1},\lambda_1\cdots,\lambda_{\alpha})-
P_t^{[\alpha]}(t;\lambda_0,\lambda_1,\cdots,\lambda_{\alpha})
P^{[\alpha]}(t;\lambda_{\alpha+1},\lambda_1\cdots,\lambda_{\alpha})}
{\lambda_0-\lambda_{\alpha+1}}~.
\nonumber
\label{A1}
\eea
Then they are given by
\beq
P^{[\alpha]}(t;\lambda_0,\lambda_1,\cdots,\lambda_{\alpha-1},\lambda_\alpha)=
c(t;\lambda_1,\cdots,\lambda_{\alpha-1})\,
\frac{\det_{i,j} P^{(i)}(t;\lambda_j)}
{
\prod_{i<j} (\l_i-\lambda_j)
}
\label{A2}
\eeq
where 
$P^{(i)}(t;\lambda)=\frac{\partial^i}{\partial t^i}P^{[0]}(t;\lambda)$,
and
$c(t;\lambda_1,\cdots,\lambda_{\alpha-1})$ is a function in $t$ 
and in $\{\lambda_1,\cdots,\lambda_{\alpha-1}\}$.

\noi
{\bf Proof} : We prove it by induction. Suppose (\ref{A2}) holds for 
an $\alpha$. Let $P^{(i)}(\l)\equiv
\frac{\partial^i}{\partial t^i}P^{[0]}(t;\l)$
(the argument $t$ is for convenience suppressed below).
Consider the numerator in (A.1),
\bea
\!\!\!\!&\!\!\!\!&\!\!\!
c(\lambda_1,\cdots,\lambda_{\alpha-1})
\left|
\ba{ccccc}
P & P' & \cdots & P^{(\a-1)} & P^{(\a)}(\l_0)\\
P & P' & \cdots & P^{(\a-1)} & P^{(\a)}(\l_1)\\
\vdots & \vdots & \cdots & \vdots & \vdots \\
P & P' & \cdots & P^{(\a-1)} & P^{(\a)}(\l_\a)
\ea
\right|
\cdot
\left(
c(\lambda_1,\cdots,\lambda_{\alpha-1})
\left|
\ba{ccccc}
P & P' & \cdots & P^{(\a-1)} & P^{(\a)}(\l_{\alpha+1})\\
P & P' & \cdots & P^{(\a-1)} & P^{(\a)}(\l_1)\\
\vdots & \vdots & \cdots & \vdots & \vdots \\
P & P' & \cdots & P^{(\a-1)} & P^{(\a)}(\l_\a)
\ea
\right|\right)' \nonumber \\
\!\!\!\!&\!\!\!\!&\!\!\!\ \ - (\l_0\leftrightarrow\l_{\a+1})\nonumber\\
\!\!\!\!&\!\!\!\!&\!\!\!=
c(\lambda_1,\cdots,\lambda_{\alpha-1})^2
\left|
\ba{ccccc}
P & P' & \cdots & P^{(\a-1)} & P^{(\a)}(\l_0)\\
P & P' & \cdots & P^{(\a-1)} & P^{(\a)}(\l_1)\\
\vdots & \vdots & \cdots & \vdots & \vdots \\
P & P' & \cdots & P^{(\a-1)} & P^{(\a)}(\l_\a)
\ea
\right|
\cdot
\left|
\ba{ccccc}
P & P' & \cdots & P^{(\a-1)} & P^{(\a+1)}(\l_{\alpha+1})\\
P & P' & \cdots & P^{(\a-1)} & P^{(\a+1)}(\l_1)\\
\vdots & \vdots & \cdots & \vdots & \vdots \\
P & P' & \cdots & P^{(\a-1)} & P^{(\a+1)}(\l_\a)
\ea
\right| \nonumber\\
\!\!\!\!&\!\!\!\!&\!\!\! \ \ - (\l_0\leftrightarrow\l_{\a+1})\nonumber\\
\!\!\!\!&\!\!\!\!&\!\!\!\equiv c(\lambda_1,\cdots,\lambda_{\alpha-1})^2
\,
D(\lambda_0,\lambda_1,\cdots,\lambda_{\alpha-1},\lambda_\alpha)~.
\eea
The above identity holds because the $t-$derivatives acting on
$c(\lambda_1,\cdots,\lambda_{\alpha-1})$ produce terms 
independent of
$\l_0$ and $\l_{\alpha+1}$ and are thus cancelled
by the subtraction interchanging $\l_0 \leftrightarrow
\l_{\alpha+1}$, 
and those acting on the columns
$$
\left(
\ba{c}
P\\
\vdots\\
P 
\ea
\right)
,
\left(
\ba{c}
P'\\
\vdots\\
P' 
\ea
\right)
, \ldots , 
\left(
\ba{c}
P^{(\a-1)}\\
\vdots\\
P^{(\a-1)} 
\ea
\right)
$$ 
vanish by themselves (the same columns are already present in the
determinant).

\noi
By definition, 
$D(\lambda_0,\lambda_1,\cdots,\lambda_{\alpha-1},\lambda_\alpha) $
enjoys the following properties:
\begin{itemize}
\item
It is a sum of monomials of the form
\beq
\pm\,P^{(\#)}(\l_0)\,P^{(\#)}(\l_1)\,P^{(\#)}(\l_1)\cdots
P^{(\#)}(\l_\a)\,P^{(\#)}(\l_\a)\,P^{(\#)}(\l_{\a+1})~,
\eeq
with (total \# of derivatives) $= 2(0+1+ \cdots + \alpha)+1=\a^2 + \a+1$.
\item
It vanishes if any two $\l$'s are coincident.
\item
It is antisymmetric under $\l_0 \leftrightarrow \l_{\alpha+1}$.
\item
It is completely symmetric under $\l_i \leftrightarrow \l_j$,
$1\leq i,j \leq \alpha$.
\end{itemize}
The only function which satisfies these properties is,
up to a constant, 
\beq
\left|
\ba{cccc}
P & P' & \cdots & P^{(\a-1)}(\l_1)\\
\vdots & \vdots  & \cdots & \vdots \\
P & P' & \cdots & P^{(\a-1)}(\l_\a)
\ea
\right|\cdot
\left|
\ba{cccccc}
P & P' & \cdots & P^{(\a-1)} & P^{(\a)}& P^{(\a+1)}(\l_0)\\
P & P' & \cdots & P^{(\a-1)} & P^{(\a)}& P^{(\a+1)}(\l_1)\\
\vdots & \vdots  & \cdots & \vdots & \vdots& \vdots  \\
P & P' & \cdots & P^{(\a-1)} & P^{(\a)}& P^{(\a+1)}(\l_\a)\\
P & P' & \cdots & P^{(\a-1)} & P^{(\a)}& P^{(\a+1)}(\l_{\a+1})
\ea
\right| ~.
\label{num}
\eeq
On the other hand, the denominator in (A.1) is given by
\beq
(\l_0-\l_{\a+1})
\prod_{0\leq i,j\leq \a} (\l_i-\l_j)
\prod_{1\leq i,j\leq \a+1} (\l_i-\l_j)
=
\prod_{1\leq i,j\leq \a} (\l_i-\l_j)
\prod_{0\leq i,j\leq \a+1} (\l_i-\l_j) ~.
\label{den}
\eeq
Since 
the first factors in (\ref{num}) and (\ref{den}) as well as
$c(\lambda_1,\cdots,\lambda_{\alpha-1})^2$ are
independent of $\l_0$ and $\l_{\a+1}$, 
it can be absorbed into the redefinition of $c(\l_1,\cdots,\l_\a)$.
Then we recover (\ref{A2}) for $\alpha+1$. QED.

\noi
Now we replace 
$\lambda_i\to{\zeta_i^2}/{N^2}$ and
$P^{[0]}(t,\l)$ by its microscopic limit,
$P^{[0]}(t;{\zeta^2}/{N^2})\to J_0(u(t)\zeta)$.
Then we can inductively prove that 
its $t$-derivatives are expressed as
\beq
P^{(i)}(t;\frac{\zeta^2}{N^2})
\to \frac{d^i}{dt^i}J_0(u(t)\zeta)
=\sum_{k=0}^i d_{i,k}(t)\,\zeta^k\,J_{k}(u(t)\zeta) ~.
\eeq
Once it is substituted inside the determinant
$\det\,P^{(i)}(\l_j)$, 
only the top term proportional to $\zeta^i\,J_{i}(u(t)\zeta)$
contributes. Thus the determinant in (\ref{A2}) is replaced by
\beq
\left(\prod_{i=0}^{\a+1}d_{i,i}(t)\right)\,\det_{0\leq i,j\leq \alpha+1} 
\zeta_j^i\,J_{i}(u(t)\zeta_j) ~.
\eeq
In order to construct the kernel,
we make an analytical continuation of $(\zeta_1,\cdots,\zeta_\a)$ 
to imaginary $(i\mu_1,\cdots,i\mu_\a)$, denote the rest as
$\zeta_0=\zeta,\ \zeta_{\a+1}=\zeta'$, and
multiply the whole expression (\ref{A2}) at $t=1$ by
the factors from the measure and the Jacobian,
$\sqrt{|\zeta\,\zeta'|
\prod_{f=1}^\a (\zeta^2+\mu^2_f)({\zeta'}^2+\mu^2_f)}$.
Then we obtain
\beq
K^{(\alpha)}(\zeta,\zeta';\{\mu_f\})=C(\{\mu_f\})\,
{\sqrt{|\zeta\,\zeta'|}}
\frac{
\left|
\begin{array}{cccc}
J_0(2\pi\rho(0)\zeta)  &\zeta\, J_1(2\pi\rho(0)\zeta) & \cdots 
& \zeta^{\alpha+1} J_{\alpha+1}(2\pi\rho(0)\zeta) \\
J_0(2\pi\rho(0)\zeta')  & \zeta'\, J_1(2\pi\rho(0)\zeta') & \cdots & 
{\zeta'}^{\alpha+1}J_{\alpha+1}(2\pi\rho(0)\zeta') \\
I_0(2\pi\rho(0)\mu_1)  &-\mu_1\, I_1(2\pi\rho(0)\mu_1) & \cdots 
&(-\mu_1)^{\alpha+1} I_{\alpha+1}(2\pi\rho(0)\mu_1) \\
\vdots    & \vdots   & \cdots           & \vdots \\
I_0(2\pi\rho(0)\mu_\alpha)  &-\mu_\alpha\, I_1(2\pi\rho(0)\mu_\alpha) 
& \cdots 
&(-\mu_{\alpha})^{\alpha+1} I_{\alpha+1}(2\pi\rho(0)\mu_\alpha) 
\end{array}
\right|
}{(\zeta^2-{\zeta'}^2)
\prod_{f = 1}^{\alpha}
\sqrt{
\left( {{\zeta }^2} + {{{{\mu }_f}}^2} \right) \,
( {{\zeta' }^2} + {{{{\mu }_f}}^2} ) 
}} ,
\label{generalk}
\eeq
\beq
\rho^{(\alpha)}_S (\zeta ; \{\mu_f\})=
{{\cal C}(\{\mu_f\})}\,
{|\zeta|}
\frac{
\left|
\begin{array}{ccccc}
\zeta^{-1}J_{-1}(\zeta)  & J_0(\zeta) & \cdots 
& \zeta^{\alpha-1}J_{\alpha-1}(\zeta) 
& \zeta^{\alpha}J_{\alpha}(\zeta) \\
J_0(\zeta)  & \zeta\,J_1(\zeta) & \cdots  
& {\zeta}^{\alpha}J_{\alpha}(\zeta) 
& {\zeta}^{\alpha+1}J_{\alpha+1}(\zeta) \\
I_0(\mu_1)  &-\mu_1\, I_1(\mu_1) & \cdots 
& (-\mu_1)^{\alpha}I_{\alpha}(\mu_1) 
& (-\mu_1)^{\alpha+1}I_{\alpha+1}(\mu_1) \\
\vdots    &  \vdots    & \cdots    &  \vdots       &  \vdots  \\
I_0(\mu_\alpha)  &-\mu_\alpha\, I_1(\mu_\alpha) & \cdots 
& (-\mu_{\alpha})^{\alpha}I_{\alpha}(\mu_\alpha) 
& (-\mu_{\alpha})^{\alpha+1} I_{\alpha+1}(\mu_\alpha) 
\end{array}
\right|
}{\prod_{f = 1}^{\alpha}
( \zeta^2 + \mu_f^2 )} ~.
\label{generalrho}
\eeq
The constant ${\cal C}(\{\mu_f\})=
(2\pi\rho(0))^{{(1-\frac{\a}{2})(1+\a)}}\,C(\{\mu_f\})$ 
is determined to be
$-(2\det{\cal M})^{-1}$ by 
requiring the matching between the 
$\zeta\rightarrow\infty$ limit of the microscopic density
(\ref{generalrho})
and the macroscopic density:
\bea
\rho^{(\alpha)}_S (\zeta\rightarrow\infty ; \{\mu_f\}) &\to&
{{\cal C}(\{\mu_f\})}\,
{|\zeta|}
\frac{
\det{\cal M}\,
\left|
\begin{array}{cc}
\zeta^{\a-1} J_{\alpha-1}(\zeta) & \zeta^{\a} J_{\alpha}(\zeta) \\
\zeta^{\a}J_{\alpha}(\zeta) & \zeta^{\a+1}J_{\alpha+1}(\zeta)
\end{array}
\right|}{\zeta^{2\a}}\cr
&\sim& -{{\cal C}(\{\mu_f\})} \frac{2}{\pi}\det{\cal M} ~.
\eea
By our normalization (\ref{mrhodef}), this should equal $1/\pi$.
Substituting the resulting ${\cal C}$ 
back to (\ref{generalk}) and (\ref{generalrho}),
we establish the result announced in section 2.3.

\end{document}